\documentclass[a4paper]{jpconf}
\usepackage{graphicx}

\newcommand{\be}{\begin{equation}}
\newcommand{\ee}{\end{equation}}
\newcommand{\bea}{\begin{eqnarray}}
\newcommand{\eea}{\end{eqnarray}}

\begin{document}

\title{Astrophysical neutrinos: theory}

\author{Ofelia Pisanti}

\address{Dipartimento di Fisica E. Pancini, Universit\`a di Napoli Federico II, and INFN, Sezione di Napoli, Via Cintia, I-80126 Napoli, Italy}

\ead{pisanti@na.infn.it}

\begin{abstract}
In the era of multi-messenger astronomy, neutrinos are among the most important astronomical messengers, due to their interaction properties. In these lessons I briefly review the main issues concerning the theory on astrophysical neutrinos.
\end{abstract}

\section{Introduction}

Unlike high-energy photons and baryons, high-energy neutrinos travel through the cosmos unmodified (except for redshift energy losses and flavour oscillations) and without significant deflections by magnetic fields.  They thus give us a probe into the high energy-phenomena of the universe, allowing us to test particle interactions and fundamental laws. However, the reason for their strength as cosmic messengers, that is a very tiny interaction with matter, is also the source of the difficulties in detecting them on Earth.

The first detection of extra-solar neutrinos dates back to the explosion of the supernova SN1987A in the Large Magellanic Cloud and, though their energy was only at the MeV scale, their detection enabled us to confirm the basic theoretical picture of the evolution of massive stars.

Then, in 2013 IceCube reported the first detection of extraterrestrial neutrinos with energies between 10 TeV and 2 PeV \cite{Aartsen:2013jdh}. After this discovery, IceCube found additional events, whose angular and energy distribution is consistent with an extra-galactic origin \cite{Kopper:2017zzm}.

So, neutrino astronomy seems to be now possible, but it also appears increasingly evident that the best way to constrain possible scenarios of cosmic neutrino production is to cross-check the information coming from their detection with the one provided by experiments on different cosmic messengers, as charged Cosmic Rays (CRs), photons or gravitational waves (GWs).

Historically, the first extra-solar multi-messenger detection happened precisely with SN1987A, because a large spectrum of electromagnetic signals and neutrinos were registered by Kamiokande II \cite{Hirata:1987hu}, the Irvine-Michigan-Brookhaven Experiment (IBM) \cite{Bionta:1987qt} and Baksan neutrino observatories \cite{Alekseev:1987ej}.

In these lectures, which cannot of course be exhaustive due to time constraints, I will review the most important theoretical concepts on astrophysical neutrinos, putting the main emphasis on the relation between neutrinos and other cosmic messengers. Following this approach, I will start from a brief summary of data on charged CRs and neutrinos. Then, I will discuss multi-messenger relations between neutrinos, CRs and photons, which will be the theoretical basis for describing neutrino production in sources and along their path to Earth. Finally, I will conclude presenting some hints connected to neutrinos from sources of gravitational waves and multi-messenger events.

\section{A unified picture of different messengers}

As stressed in \cite{Mannheim:1998wp}, ``... cosmic proton accelerators produce cosmic rays, gamma rays, and neutrinos with comparable luminosities...". This circumstance is compatible with the fact that all different messengers are connected among them by the underlying physics. In particular, according to the present paradigm, CRs are at the beginning of the production chain; for example, a fraction of the gravitational energy of relativistic particles in shocks near Neutron Stars (NSs) or Black Holes (BHs) can be transformed into the acceleration of protons, which produce $\gamma$'s and $\nu$'s before arriving to Earth as CRs. So, I start reviewing what we know about charged CRs.

\subsection{Charged CRs}

\begin{figure}[t]
\includegraphics[width=.5\textwidth]{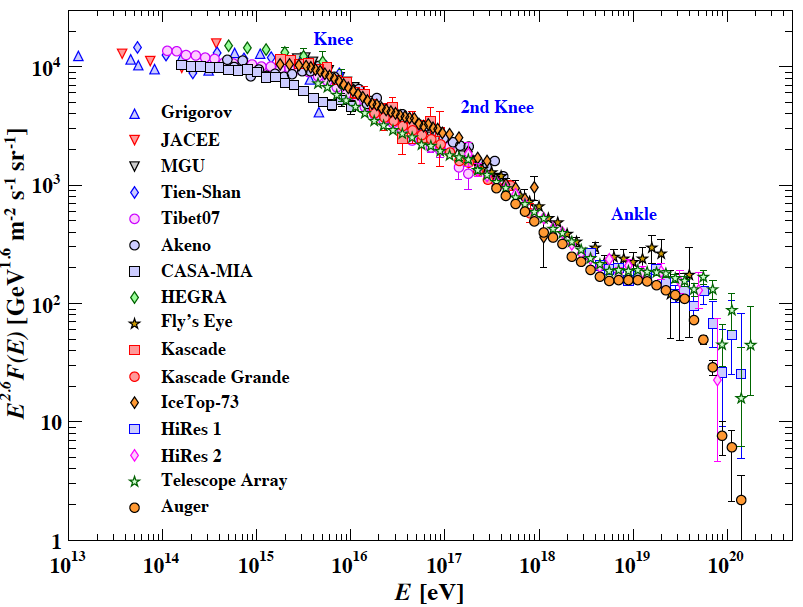}\hspace{2pc}%
\begin{minipage}[b]{.44\textwidth}\caption{\label{CRspectrum}All-particle CR spectrum as a function of the energy-per-nucleus from air-shower measurements \cite{Patrignani:2016xqp}.}
\end{minipage}
\end{figure}

As well known, CRs spectrum is represented by a broken power law spanning over more than 10 orders of magnitude (see figure \ref{CRspectrum}). In the so-called bottom-up scenario CRs below the knee are likely produced by supernovae, but those at higher energies have a less certain origin. The production/acceleration of CRs could be achieved in a plethora of sources, like Active Galactic Nuclei (AGNs), Gamma Ray Bursts (GRBs), magnetars (NSs with petagauss surface magnetic fields), accretion shocks around cluster of galaxies, etc.. In particular, Ultra-High Energy CRs (UHECRs) come almost certainly from extragalactic sources, because they cannot be confined by galactic magnetic fields.

Competing to the bottom-up is the top-down scenario, where CRs originate from the decay/annihilation of particles beyond the standard model. While highly model dependent, this kind of models seem to be not favoured by the lack of detected photons and neutrinos in the ultra-high energy range of the CR spectrum \cite{Mariazzi:2018yap}.

While pretty well known at low energy, mass composition of UHECRs is still debated. On one side, the Paul Auger Observatory (PAO) observes a cutoff at $4\cdot 10^{19}$ eV and a mixed mass composition (protons at energies below $10^{18}$ eV, nuclei at high energy) \cite{Bellido:2017cgf}. On the other side, the Telescope Array experiment (TA) finds a cutoff at $5.4\cdot 10^{19}$ eV and a light mass composition \cite{Ikeda:2017qvn}. Note, however, that the mean value and root-mean-square of the measured atmospheric depth at maximum, $X_{max}$, from the two experiments are consistent within systematic uncertainties \cite{deSouza:2017wgx}.

Motivated by these two somehow conflicting results, two main scenarios have been considered\footnote{Other interpretations of data exist, see for example in this proceedings the lectures by Dmitry Semikoz.}. In the first scenario (PAO data) both the ankle in the spectrum and the high energy suppression are due to rigidity effects ($E_{max} (Z)= Z\, E_{max,p}$) and nuclei photo-disintegration. In the second one (TA data), the so called {\it dip} model \cite{Berezinsky:1988wi,Berezinsky:2002nc,Berezinsky:2005cq}, the ankle is an effect of the pair production by protons on the Cosmic Microwave Background (CMB) and the suppression is due to proton interactions with the same target (GZK effect, see later).

The deflection of UHECR by extragalactic and galactic magnetic field can be low enough to use them for finding anisotropies. Both PAO and TA observe the GZK suppression, which would imply that the sources have distances shorter than $\sim 200$ Mpc. Indications of an anisotropic distributions of sources were found; in particular, isotropy is disfavored at the 4-$\sigma$ level when catalogs of nearby starburst galaxies sources are considered \cite{Mariazzi:2018yap}.

\subsection{Neutrinos}

At intermediate energy (1 TeV $<E<$ 10 PeV) neutrino events are detected at Neutrino Telescopes, like IceCube or ANTARES, with methods aiming to reject the atmospheric muon background, as for example the use of muon vetoes or by combining event angular estimates with energy-related variables.  Icecube has observed about 80 neutrino events with deposited energies from 20 TeV to 2 PeV \cite{Kopper:2017zzm}. Analysis of data gives a power law with spectral index $\gamma = 2.92_{-0.29}^{+0.33}$. On the other end, in nine years ANTARES observes a mild excess of high-energy events over the expected background, which rejects the null cosmic flux assumption only at 1.6-$\sigma$ \cite{Albert:2017nsd}.

\begin{figure}[t]
\includegraphics[width=.5\textwidth]{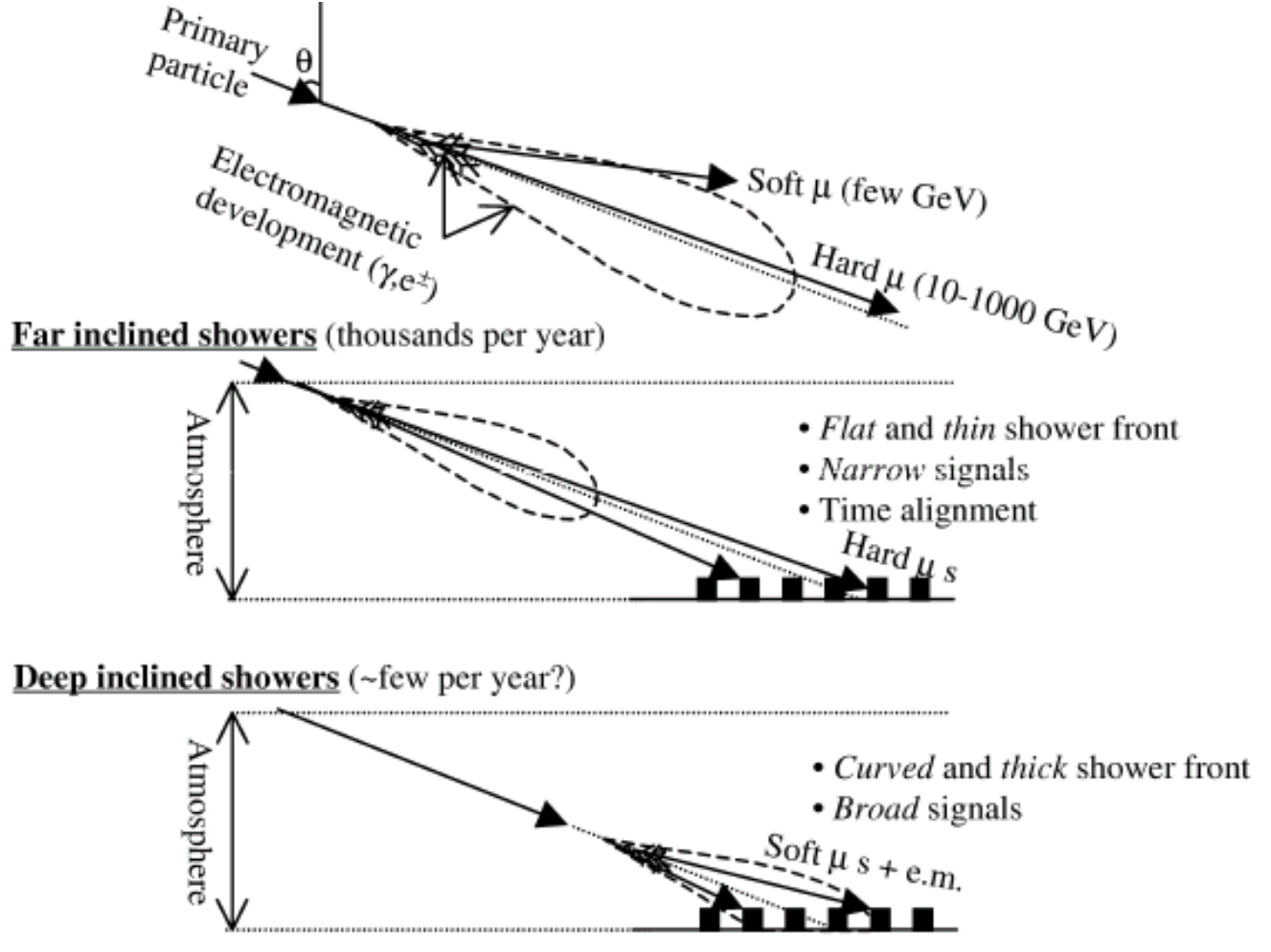}\hspace{2pc}%
\begin{minipage}[b]{.44\textwidth}\caption{\label{inclined_showers}Inclined showers induced by a charged CR and a neutrino.}
\end{minipage}
\end{figure}
At higher energy, both PAO and TA are hybrid experiments with a surface array of Cherenkov/scintillators detectors combined with fluorescence detectors, sensible to CRs above $\sim 10^{18}$ eV. At extensive air shower array experiments, vertical neutrino induced showers cannot be distinguished from ordinary CR showers. But in very inclined showers it is possible to identify different features for the different primaries, see figure \ref{inclined_showers}. In fact, the electromagnetic component of a shower induced by a charged CR gets absorbed due to the large grammage of atmosphere from the first interaction point to the ground. As a consequence, the shower front at ground level is dominated by muons that induce sharp time traces in the water-Cherenkov stations. On the contrary, showers induced by downward-going neutrinos at large zenith angles can start their development deep in the atmosphere, producing traces that spread over longer times. In any case, at the moment the two experiments give only bounds on cosmic neutrino spectrum \cite{Zas:2017xdj,Rubtsov:2017lhs}.

\begin{figure}[t]
\includegraphics[width=.6\textwidth]{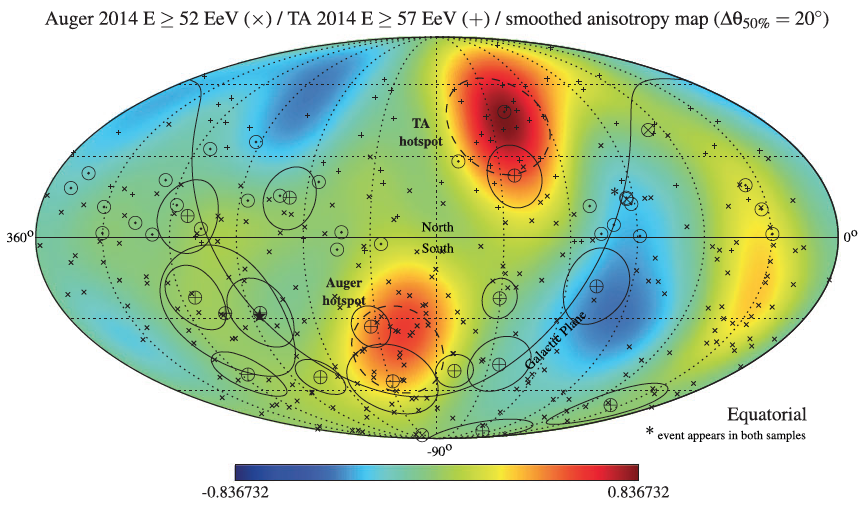}\hspace{2pc}%
\begin{minipage}[b]{.34\textwidth}\caption{\label{sky_map}Arrival directions of neutrinos and UHE CRs \cite{Aartsen:2015dml}. The excess regions found by PAO and TA are highlighted.}
\end{minipage}
\end{figure}
A combined analysis using the CR events observed by PAO above 52 EeV \cite{PierreAuger:2014yba} and TA above 57 EeV \cite{Abbasi:2014lda} and the neutrino events detected by IceCube \cite{Aartsen:2016xlq} was carried out in \cite{Aartsen:2015dml}. The anisotropy map by IceCube reported in figure \ref{sky_map} shows two small-scale excess regions in the Northern and Southern Hemisphere that coincide with the excess regions reported by PAO and TA, but no noticeable clustering of high-energy neutrino events in the direction of these hotspots are present. This does not necessarily rule out a correlation between CR and neutrino sources, because neutrinos can come from distances up to the Hubble horizon while CRs above $\sim 50$ EeV only from local sources up to 200 Mpc. Comparing these distances, one can estimate that ony 5\% of $\nu$ should correlate with CRs, corresponding to two events over 45 shown on the map \cite{Ahlers:2017wkk}.

\section{Multi-messenger relations}
\label{mmsection}

\begin{figure}[b]
\begin{center}
\begin{minipage}{.8\textwidth}
\includegraphics[width=.9\textwidth]{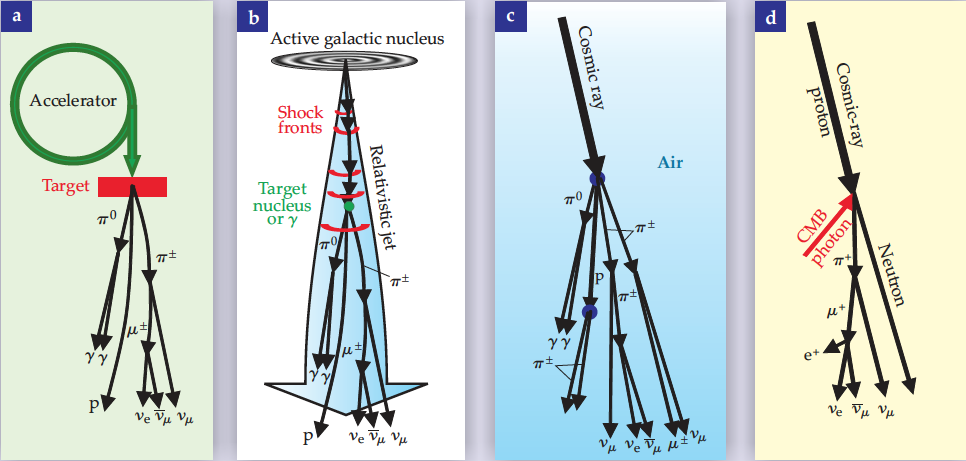}
\end{minipage}
\end{center}
\caption{\label{nuint}Interactions that produce neutrinos in a) accelerators, b) cosmic sources, c) atmosphere, d) along path to Earth (figure from \cite{Halzen:2008zz}).}
\end{figure}
In this section I review the physics mechanisms by which neutrinos are produced in the cosmo. This applies in astrophysical sources or along the CR path to Earth or in their interaction with our atmosphere and, of course, also when neutrinos are produced at accelerators, see figure \ref{nuint}. In the following, I will combine together neutrino and anti-neutrino fluxes.

\begin{figure}[t]
\includegraphics[width=.5\textwidth]{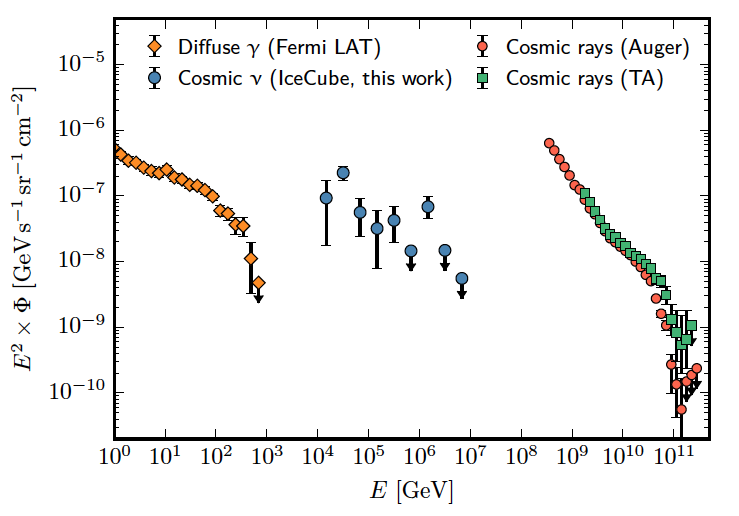}\hspace{2pc}%
\begin{minipage}[b]{.44\textwidth}\caption{\label{messengers}Neutrino flux from Icecube \cite{Mohrmann:2015lag}, compared to $\gamma$ flux from Fermi LAT \cite{Ackermann:2014usa} and CR flux from TA \cite{AbuZayyad:2012ru} and PAO \cite{ThePierreAuger:2013eja}.}
\end{minipage}
\end{figure}

The fact that the total energy density of neutrinos is similar to that of gamma rays, see figure \ref{messengers}, suggests some connection between them (and with CRs). And in fact the main mechanism at the origin of both is the production of pions in the interactions of CRs with ambient matter, which later decay to $\gamma$'s ($\pi^0$) and $\nu$'s ($\pi^\pm$).

In Pion Photo-production (PP),
\be
p+\gamma\rightarrow p+\pi^0, ~~~p+\gamma\rightarrow n+\pi^+,
\ee
charged and neutral pions are produced resonantly (with 2/3 and 1/3 probabilities, respectively) and non-resonantly (changing the previous probabilities to 1/2 and 1/2) and the inelasticity, that is the energy released to the secondary pion, is $\kappa_\pi\simeq 0.2$ \cite{Kelner:2006tc}.

In Hadronic Collisions (HC),
\be
p+p\rightarrow \pi+X,
\ee
positive, negative, and neutral pions are produced with equal probability, 1/3 and inelasticity is $\kappa_\pi\simeq 0.5$ \cite{Kelner:2006tc}.

Then, in the charged pion decay,
\be
\pi^\pm\rightarrow \mu^\pm +\nu_\mu\rightarrow e^\pm + \nu_e + \nu_\mu,
\label{pipmdecay}
\ee
the energy of each neutrino can be approximated as $\sim$25\% of pion energy, $\kappa_\nu\simeq 0.25$, while in the neutral pion decay,
\be
\pi^0\rightarrow \gamma + \gamma,
\label{pi0decay}
\ee
the energy of each photon is $\sim 50$\% of the pion energy, $\kappa_\gamma\simeq 0.5$.

Putting all this together we have the following:
\begin{itemize}
\item
by defining the quantity $K_\pi=1(2)$ for PP(HC), we can write down the relation between the number of charged and neutral pions and the probability of production of the charged pions, 
\bea
N_{\pi^\pm} &=& K_\pi N_{\pi^0}, \\
P_{\pi^\pm} &=& \frac{K_\pi}{1+K_\pi}; \label{piprob}
\eea
\item
since in PP $\kappa_\pi\simeq 0.2$ and $\kappa_\nu\simeq 0.25$, neutrinos have $\sim 5$\% of the energy of the original proton; then
\be
E_\nu \simeq 0.05\, E_p \equiv 0.05\, 10^{17} E_{p,17} = 5\, \mathrm{PeV}\, \frac{\epsilon_{p,17}}{1+z},
\ee
where we have taken into account the proton redshift energy loss from the neutrino production point, $\epsilon_{p,17}=(1+z)\, E_{p,17}$. This relation suggests that neutrinos around PeV could be produced by pion photo-production of protons with energies close to the second knee, $\sim 10^{17}$ eV;
\item
since in PP $\kappa_\pi\simeq 0.2$ and $\kappa_\gamma\simeq 0.5$, neutrinos have $\sim 10$\% of the energy of the original proton.
\end{itemize}

Using equation (\ref{piprob}), the pion production rate, that is the number of charged pions per unit energy and time, results proportional to the corresponding CR production rate,
\be
Q_{\pi^\pm} (E_\pi) \equiv \frac{dN_{\pi^\pm}}{dt\, dE}  (E_\pi) = \frac{K_\pi}{1+K_\pi}\, \left[ Q_N (E_N) \right]_{E_N=E_\pi/\kappa_\pi}.
\label{Qpip}
\ee

Actually, efficient acceleration of cosmic rays would imply that any loss mechanism that reduces the acceleration time, like pion production, should be suppressed. Optically thick sources, where the ambient matter is very dense, would be efficient neutrino emitters while neutrino production would be marginal in optically thin sources. In order to take into account this difference, one introduces a factor $f_\pi$, characterizing thin ($f_\pi\ll 1$) or thick ($f_\pi\sim 1$) sources, defined as in the following equation:
\be
E_\pi^2\, Q_{\pi^\pm} \simeq f_\pi\, \frac{K_\pi}{1+K_\pi}\, \left[ E_N^2\, Q_N (E_N) \right]_{E_N=E_\pi/\kappa_\pi}.
\label{Qpi}
\ee

We now connect pion production rates to photon and neutrino ones using the information from particle physics. Since we have 2 $\nu_\mu$ for each charged pion, see equation (\ref{pipmdecay}), the total number of charged pions with energy between $E_1$ and $E_2$ is
\be
N_{\pi^\pm} = \frac{1}{2}\, \int_{\kappa_\nu E_1}^{\kappa_\nu E_2} \frac{dN_{\nu_\mu}}{dE_\nu}\, dE_\nu.
\ee
Similarly, since we have 2 $\gamma$ for each $\pi_0$, see equation (\ref{pi0decay}), the total number of $\pi_0$ with energy between $E_1$ and $E_2$ is
\be
N_{\pi^0} = \frac{1}{2}\, \int_{\kappa_\gamma E_1}^{\kappa_\gamma E_2} \frac{dN_\gamma}{dE_\gamma}\, dE_\gamma.
\ee
These equations, upon derivation with respect to $E_2$, gives
\be
\frac{\kappa_\nu}{2}\, \left. \frac{dN_{\nu_\mu}}{dE_\nu}\right|_{E_\nu=\kappa_\nu E} = \left. \frac{dN_{\pi^\pm}}{dE_\pi}\right|_{E}
\label{numufromchpi}
\ee
and
\be
\frac{\kappa_\gamma}{2}\, \left. \frac{dN_\gamma}{dE_\gamma}\right|_{E_\gamma=\kappa_\gamma E} = \left. \frac{dN_{\pi^0}}{dE_\pi}\right|_{E}.
\ee
Then, going to the production rates, we have
\bea
Q_{\nu_\mu} (E_\nu) &=& \frac{2}{\kappa_\nu}\, Q_{\pi^\pm}  \left( \frac{E_\nu}{\kappa_\nu} \right) \simeq 8\, Q_{\pi^\pm}(4\,E_\nu), \label{Qnumu} \\
Q_{\nu_e} (E_\nu) &=& \frac{1}{\kappa_\nu}\, Q_{\pi^\pm} \left( \frac{E_\nu}{\kappa_\nu} \right) \simeq 4\, Q_{\pi^\pm}(4\,E_\nu), \label{Qnue} \\
Q_\gamma (E_\gamma) &=& \frac{2}{\kappa_\gamma}\, Q_{\pi^0} \left( \frac{E_\gamma}{\kappa_\gamma} \right) \simeq 4\, Q_{\pi^0}(2\,E_\nu), \label{Qgamma}
\eea
where equation (\ref{Qnue}) comes from the ratio $N_{\nu_\mu}/N_{\nu_e}=2$ in the charged pion decay, together with
\be
Q_{\pi^\pm} (E_\pi) = K_\pi\, Q_{\pi^0} (E_\pi).
\label{Qchpi}
\ee

Putting together equations (\ref{Qnumu}) and (\ref{Qnue}) one gets (assuming that oscillations produce 1:1:1 flavour ratios)
\be
\frac{1}{3}\, \sum_\alpha Q_{\nu_\alpha} (E_\nu) \simeq \frac{1}{3}\, (Q_{\nu_\mu} (E_\nu) + Q_{\nu_e}(E_\nu) ) \simeq 4\, Q_{\pi^\pm} (4\, E_\nu)
\label{starting1}
\ee
and, by multiplying by $E_\nu$,
\be
\frac{1}{3}\, \sum_\alpha E_\nu\, Q_{\nu_\alpha} (E_\nu) \simeq 4\, E_\nu\, Q_{\pi^\pm} (4\, E_\nu) = \left[ E_\pi\, Q_{\pi^\pm} (E_\pi) \right]_{E_\pi = E_\nu/\kappa_\nu}.
\label{starting2}
\ee

Equations (\ref{starting1}) and (\ref{starting2}) are the starting points for obtaining two very useful relations. First of all, using equations (\ref{Qpi}) and (\ref{starting2}) we can connect neutrino production rate with the parent CR one,
\be
\frac{1}{3}\, \sum_\alpha E_\nu^2\, Q_{\nu_\alpha} (E_\nu) \simeq \kappa_\nu\,f_\pi\, \frac{K_\pi}{1+K_\pi}\, \left[ E_N^2\, Q_N (E_N) \right]_{E_N=E_\nu/(\kappa_\pi\,\kappa_\nu)}.
\label{multi1}
\ee
Then, noting that equations (\ref{Qnumu}) and (\ref{Qnue}) can be written also as, see equation (\ref{Qchpi}),
\bea
Q_{\nu_\mu} (E_\nu) &=& \frac{2\, K_\pi}{\kappa_\nu}\, Q_{\pi^0}  \left( \frac{E_\nu}{\kappa_\nu} \right) \\
Q_{\nu_e} (E_\nu) &=& \frac{K_\pi}{\kappa_\nu}\, Q_{\pi^0} \left( \frac{E_\nu}{\kappa_\nu} \right)
\eea
we arrive from equation (\ref{starting1}) to
\be
\frac{1}{3}\, \sum_\alpha Q_{\nu_\alpha} (E_\nu) \simeq \frac{K_\pi}{\kappa_\nu}\, Q_{\pi^0} \left( \frac{E_\nu}{\kappa_\nu} \right) \simeq \frac{\kappa_\gamma\, K_\pi}{2\, \kappa_\nu}\, Q_\gamma\, \left( \frac{\kappa_\gamma\,E_\nu}{\kappa_\nu} \right) \simeq K_\pi\, Q_\gamma\, ( 2\,E_\nu),
\ee
and then
\be
\frac{1}{3}\, \sum_\alpha E_\nu^2\, Q_{\nu_\alpha} (E_\nu) \simeq \frac{\kappa_\nu\,K_\pi}{2\,\kappa_\gamma}\, \left[ E_\gamma^2\, Q_\gamma (E_\gamma) \right]_{E_\gamma=\kappa_\gamma\,E_\nu/\kappa_\nu} \simeq \frac{K_\pi}{4}\, \left[ E_\gamma^2\, Q_\gamma (E_\gamma) \right]_{E_\gamma=2\,E_\nu},
\label{multi2}
\ee
which is a relation somehow independent of the CR detailed model considered.

\begin{figure}[t]
\includegraphics[width=.5\textwidth]{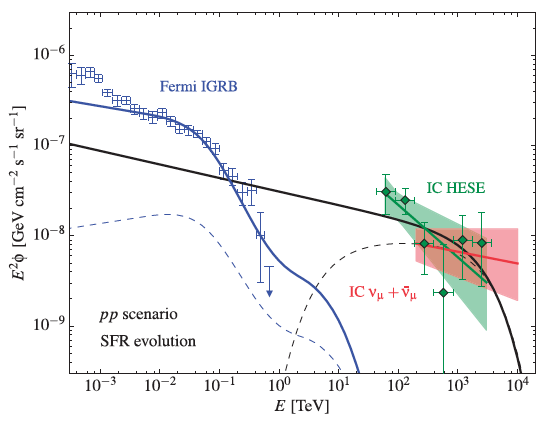}\hspace{2pc}%
\begin{minipage}[b]{.44\textwidth}\caption{\label{multi_comparison}Neutrino flux in two different models, see \cite{Ahlers:2017wkk}, (black solid and dashed lines) with the corresponding photon flux from equation (\ref{multi2}) (blue solid and dashed lines).}
\end{minipage}
\end{figure}

Figure \ref{multi_comparison} from \cite{Ahlers:2017wkk} gives an example of the application of this kind of multi-messenger relations for deriving the gamma ray spectrum associated to a given model of neutrino production.

\section{Neutrino production in sources}

I now discuss briefly neutrino production in sources, starting from the description of some bounds that one can derive on neutrino flux and then considering the effect on flavour composition at Earth due to oscillations.

\subsection{Bounds on neutrino diffuse flux}

The reasoning behind this type of bounds works as follows. CRs can be attenuated by interactions with background photons; this implies that the observed CRs come from the local (200 Mpc, $z<1$) universe. Starting from the measure of their flux one go back to the CR production rate in the whole universe by postulating the evolution of sources with redshift. Then, on the basis of relations between CR and neutrino production rates, as the ones derived in the previous section, it is possible to derive bounds on neutrino flux.

The Waxman-Bahcall bound (WB) \cite{Waxman:1998yy,Bahcall:1999yr} was derived under the following assumptions:
\begin{itemize}
\item
UHECRs are protons from optically thin sources with Fermi acceleration spectrum $dN_p/dE\!\!\!\sim\!\!\!\!{E^{-2}}$, normalized to the measured UHECR flux above EeV energies, $E^2\, dN_p/(dt\,dE) \sim 10^{44}$ erg/(Mpc$^3$ yr);
\item
all of the energy of protons is transferred to pions inside sources (this is a conservative hypothesis, since pions receive a fraction of the energy of the proton);
\item
no magnetic fields are considered.
\end{itemize}
As it is, the Waxman-Bahcall bound can be evaded for: 1) neutrinos produced by other types of interactions (like in top-down scenarios); 2) sources with high optical depth (like AGN where neutrinos are produced in the core and not in the jets).

The Mannheim-Protheroe-Rachen bound (MPR) \cite{Mannheim:1998wp} 1) relax the Waxman-Bahcall hypothesis of injection spectrum as $E^{-2}$, considering the observable spectrum of CRs, complying with existing observational limits; and 2) consider also thick sources.

However, some authors \cite{Murase:2010gj} claim that WB and MPR should be considered as ``landmarks" and not bounds, that is nominal scales instead of observational bounds, and they derive updated landmarks, motivated by recent experimental results. Their analysis is based on the following facts:
\begin{itemize}
\item
in the case of nuclei in radiation fields, the photo-disintegration process is even more important than the photo-production process;
\item
if confirmed, the heavy composition of CRs indicated by PAO data would imply that more nuclei survive photo-disintegration; this is only possible if less target photons are present in the sources and then less neutrinos would be produced by pion photo-production.
\end{itemize}
\begin{figure}[t]
\begin{minipage}{.47\textwidth}
\includegraphics[width=1.\textwidth]{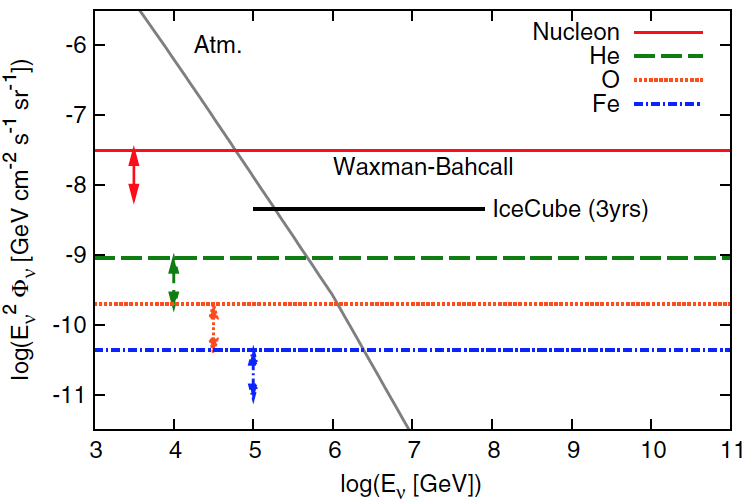}
\end{minipage}\hspace{2pc}%
\begin{minipage}{.47\textwidth}
\includegraphics[width=1.\textwidth]{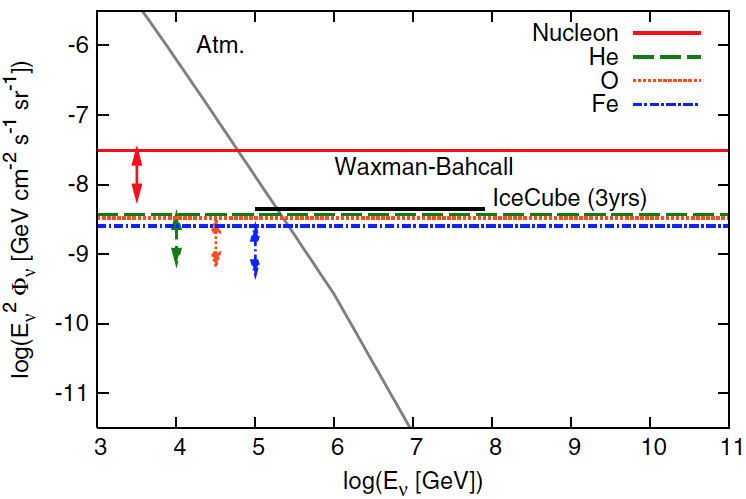}
\end{minipage} 
\caption{\label{landmark}Upper bounds for the $\nu$ background from UHECR photo-disintegration interactions in sources of different nuclei \cite{Murase:2010gj} compared with WB landmarks. Left: optical depth is used, right: effective optical depth is used. The arrows indicate the change for no redshift evolution.}
\end{figure}

The consequences of these observations are that neutrino bounds should be lower than WB or MPR. In figure \ref{landmark} the upper bounds for the $\nu$ background flux obtained for different nuclei are shown, assuming the same injection spectrum as in \cite{Waxman:1998yy,Bahcall:1999yr}, under the hypothesis that the photo-disintegration process happens mainly via the Giant Dipole Resonance. Left plot is obtained requiring optical depth less than 1, while right plot considers an effective optical depth. In both cases, small neutrino fluxes are obtained, at least one order of magnitude below the WB flux for UHE protons.

Another bound can be derived, based on the fact that IceCube has failed in identifying extra-galactic sources in the diffuse neutrino flux. If these sources, each of them with a production rate $Q_{\nu_\alpha} (E_\nu)$ for a given flavour, are distributed with a redshift density $\rho(z)$, then the sources between $z$ and $z+dz$ in the solid angle element $d\Omega$ contribute to the flux at Earth, $\phi_{\nu_\alpha} (E_\nu)$, with $\rho(z)\,Q_\nu\,d\chi\, d\Omega$ ($\chi$ = comoving distance). Then, the flux per unit solid angle of a given neutrino flavour is \cite{Ahlers:2017wkk}
\be
\phi_{\nu_\alpha} (E_\nu) = \frac{1}{4\,\pi}\, \int_0^\infty \frac{dz}{H(z)}\, \rho(z)\, Q_{\nu_\alpha} \left( (1+z)\, E_\nu \right),
\label{phinua1}
\ee
where the production rate $Q_{\nu_\alpha}$ have been calculated at the energy corresponding to the redshift $z$. Note that this corresponds to the integral on the comoving volume of individual neutrino point sources (PS) with density $\rho(z)$ and flux given by
\be
\phi_{\nu_\alpha}^{PS} (E_\nu) = \frac{(1+z)^2}{4\,\pi\, d_L^2(z)}\, Q_{\nu_\alpha} \left( (1+z)\, E_\nu \right),
\ee
as one can check by employing the expressions of comoving volume, $dV_c=r^2\,d\Omega\,dz/H(z)$ and luminosity distance, $d_L=(1+z)\,r$. If we define the spectral emission rate density, $L_{\nu_\alpha} (z,E)\equiv \rho (z)\, Q_{\nu_\alpha} (E)$, and introduce the quantity \cite{Ahlers:2014ioa}
\be
\xi_z (E) = \int_0^\infty dz\, \frac{H_0}{H(z)}\, \frac{L_\nu (z,(1+z)\,E)}{L_\nu(0,E)},
\ee
it is easy to show that equation (\ref{phinua1}) becomes
\be
\phi_{\nu_\alpha} (E_\nu) = \frac{1}{4\,\pi}\, \frac{\xi_z (E_\nu)}{H_0}\, L_{\nu_\alpha} (0,E_\nu) = \frac{1}{4\,\pi}\, \frac{\xi_z (E_\nu)}{H_0}\, \rho (0)\, Q_{\nu_\alpha} (E_\nu),
\ee
and then ($\rho_0\equiv \rho (0)$)
\be
\frac{1}{3}\, \sum_\alpha E_\nu^2\, \phi_{\nu_\alpha} (E_\nu) = \frac{1}{4\,\pi}\, \frac{\xi_z (E_\nu)}{H_0}\, \rho_0\, \frac{1}{3}\, \sum_\alpha E_\nu^2\, Q_{\nu_\alpha} (E_\nu).
\label{diffusenu}
\ee
The l.h.s. of this relation can be evaluated using the diffuse flux measured by IceCube, $E^2$ $\sum_\alpha \phi_{\nu_\alpha} \simeq 10^{-8}$ GeV cm$^{-2}$ s$^{-1}$ sr$^{-1}$ for energies in excess of $\simeq 100$ TeV \cite{Ahlers:2014ioa}. Equation (\ref{diffusenu}) can then be employed for inferring the production rate for single source \cite{Ahlers:2014ioa}
\be
\frac{1}{3}\, \sum_\alpha E_\nu^2\, Q_{\nu_\alpha} (E_\nu) \simeq 1.8\times 10^{43} \left( \frac{\xi_z}{2.4} \right)^{-1}\, \left( \frac{\rho_0}{10^{-8}\, \mathrm{Mpc}^{-3}} \right)^{-1}\, \mathrm{erg s}^{-1}.
\ee

A bound on the $\nu$ flux at Earth arises from the previous connection between neutrinos and CRs. In fact, equations (\ref{multi1}) and (\ref{diffusenu}) result in
\be
\frac{1}{3}\, \sum_\alpha E_\nu^2\, \phi_{\nu_\alpha} (E_\nu) \simeq \frac{1}{4\,\pi}\, \frac{\xi_z (E_\nu)}{H_0}\, \rho_0\, \frac{\kappa_\nu\,f_\pi\, K_\pi}{1+K_\pi}\, \left[ E_N^2\, Q_N (E_N) \right]_{E_N=E_\nu/(\kappa_\pi\,\kappa_\nu)}.
\ee
By defining $L_N (E_N) \equiv \rho_0\, E_N^2\, Q_N (E_N)$, this becomes
\be
\frac{1}{3}\, \sum_\alpha E_\nu^2\, \phi_{\nu_\alpha} (E_\nu) \simeq \frac{1}{4\,\pi}\, \frac{\xi_z (E_\nu)}{H_0}\, \frac{\kappa_\nu\,f_\pi\, K_\pi}{1+K_\pi}\, \left[ L_N (E_N) \right]_{E_N=E_\nu/(\kappa_\pi\,\kappa_\nu)}.
\ee
Assuming $L_N (E_N) \simeq (1-2)\times 10^{44}\, \mathrm{erg~Mpc}^{-3}\, \mathrm{yr}^{-1}$ \cite{Ahlers:2014ioa}, $f_\pi =1$, and $K_\pi =1,2$ for PP/HC, and a star formation rate (SFR) evolution, $\xi_z\sim 2.4$, this bound is at the level of the neutrino flux observed by IceCube.

\subsection{Flavour studies}

It is often assumed in flavour studies of neutrino flux at Earth that, due to the known pattern of neutrino oscillations, encoded in the neutrino PMNS \cite{Maki:1962mu} mixing matrix, the initial flavour composition at the source, $\phi_{\nu_e}$ : $\phi_{\nu_\mu}$ : $\phi_{\nu_\tau}=1:2:0$, becomes $\sim 1:1:1$ to Earth. This, however, is true only for the so called pion sources and instead, strictly speaking, there are other possibilities. In particular, we have
\begin{itemize}
\item
pion sources, where purely hadronic processes happen, like the ones considered in section \ref{mmsection}
\be
\phi_{\nu_e} : \phi_{\nu_\mu} : \phi_{\nu_\tau} =1:2:0;
\ee
\item
muon damped sources, where the muons may lose energy, due to strong magnetic fields, or get absorbed in matter; then the $\nu_e$ flux is depleted at the energies of interest because the length for muon energy loss is smaller than its decay length
\be
\phi_{\nu_e} : \phi_{\nu_\mu} : \phi_{\nu_\tau} =0:1:0;
\ee
\item
neutron sources (where neutrons originate in photo-dissociation of heavy nuclei) which are pure $\bar{\nu}_e$ beam
\be
\phi_{\nu_e} : \phi_{\nu_\mu} : \phi_{\nu_\tau} =1:0:0.
\ee
\end{itemize}

Since the distances to high energy neutrino sources are usually quite large compared to oscillation lengths, the terms with the mass-squared differences are averaged out and oscillation probabilities are reduced to mixings. The flux for a given flavour $\nu_\alpha$ at Earth is given, in terms of the fluxes at the source, $\phi_{\nu_\beta}^0$, by
\be
\phi_{\nu_\alpha} = \sum_\beta P_{\alpha\beta} \, \phi_{\nu_\beta}^0,
\ee
where $P_{\alpha\beta}$ are the mixing probabilities
\be
P_{\alpha\beta} = \sum_i | U_{\alpha\,i}|^2 | U_{\beta\,i}|^2.
\ee
A useful exercise is to calculate the flavour ratios at Earth when a simple phenomenological form for the probability matrix is adopted, the so-called tri-bimaximal form of $P$ \cite{Harrison:2002er},
\be
P_{TBM} = \frac{1}{18}\, \left( 
\begin{array}{ccc}
10 & 4 & 4 \\
4 & 7 & 7 \\
4 & 7 & 7 \\
\end{array}
\right).
\ee
For example, starting from the pion source case, we have
\be
\Phi_\nu = \frac{1}{18}\, \left( 
\begin{array}{ccc}
10 & 4 & 4 \\
4 & 7 & 7 \\
4 & 7 & 7 \\
\end{array}
\right)
\frac{\phi_\nu^0}{3}\, \left( 
\begin{array}{c}
1 \\
2 \\
0 \\
\end{array}
\right) = 
\frac{\phi_\nu^0}{3}\, \left( 
\begin{array}{c}
1 \\
1 \\
1 \\
\end{array}
\right),
\ee
where $\phi_\nu^0 = \phi_{\nu_e}^0+\phi_{\nu_\mu}^0+\phi_{\nu_\tau}^0$.

Authors in \cite{Pakvasa:2007dc} investigate the results one can obtain considering different event topologies at experiments. In particular, they assume that:
\begin{itemize}
\item[i)]
$\nu_\mu$ events can be identified via muons (tracks);
\item[ii)]
$\nu_e$ charged current reactions may be identifiable (showers);
\item[iii)]
at energies $> 10^6$ GeV, $\nu_\tau$ can be identified by double-bang or lollipop signatures.
\end{itemize}
They focus on two observables, the ratio of $\nu_\mu$ events to the total flux, $T\equiv \phi_{\nu_\mu}^0/\phi_\nu^0$, and of $\nu_e$ to $\nu_\tau$ events, $R\equiv \phi_{\nu_e}/\phi_{\nu_\tau}$, even if other observable can be considered, like the following two
\be
\frac{\phi_{\nu_\mu}}{\phi_{\nu_e} + \phi_{\nu_\tau}} = \frac{T}{1-T}, ~~~~~~~~ \frac{\phi_{\bar{\nu}_e}}{\phi_{tot}}.
\ee
For the previous tri-bimaximal form of $P$, the following predictions can be done, apart from small corrections (see after),
\begin{itemize}
\item
pion sources: $T=1/3$, $R=1$;
\item
muon damped sources: $T=7/18$, $R=4/7$;
\item
neutron sources: $T=2/9$, $R=5/2$.
\end{itemize}

\begin{figure}[t]
\begin{minipage}{.47\textwidth}
\includegraphics[width=1.\textwidth]{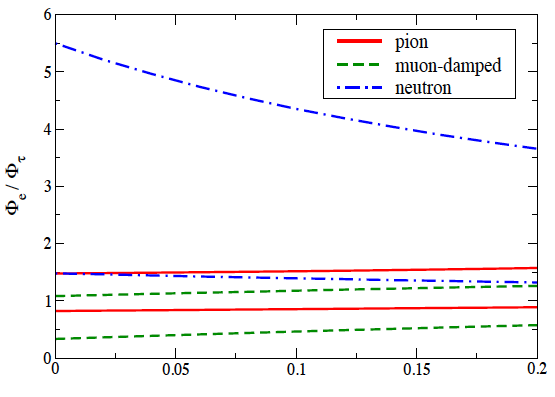}
\end{minipage}\hspace{2pc}%
\begin{minipage}{.47\textwidth}
\includegraphics[width=1.\textwidth]{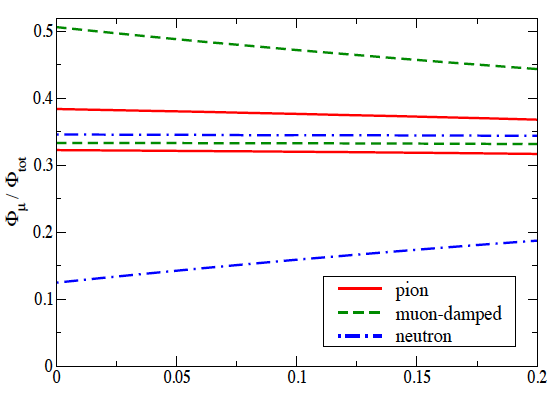}
\end{minipage} 
\caption{\label{flavratios}Minimal and maximal values of T and R (see text) obtained by varying neutrino oscillation parameters within their 3-$\sigma$ ranges \cite{Pakvasa:2007dc}. On the horizontal axis the deviation from idealized flux composition is considered.}
\end{figure}

The information coming from the measure of this kind of observables at experiments are complicated by the fact that one expects deviations from the idealized initial flavour ratios. These derive both from approximations on the inelasticity of charged pions and muons and from the assumption that the muon energy loss before decay is negligible. Other corrections come from the fact that the production and decay of kaons also produces neutrinos. Correcting for these approximations results in some difference with respect to the idealized flavour ratios, for example \cite{Lipari:2007su}
\be
\phi_{\nu_e} : \phi_{\nu_\mu} : \phi_{\nu_\tau} =1:2:0 ~~ \rightarrow ~~ \phi_{\nu_e} : \phi_{\nu_\mu} : \phi_{\nu_\tau} =1.00 : 1.85 : 0.001
\ee
Then it is necessary to estimate how flavour observables, like $T$ and $R$, change as a function of these deviations. The analysis made in \cite{Pakvasa:2007dc}, by varying oscillation parameters within their 3-$\sigma$ ranges, shows that the muon-damped and neutron sources are more sensitive to initial flavor deviations than are pion sources, see figure \ref{flavratios}.

\section{Cosmogenic neutrinos}

As previously said, efficient neutrino production would require different zones, one for the acceleration and one for conversion to neutrinos. Actually, this is realized in the case of cosmogenic $\nu$'s. These neutrinos have the additional advantage that they can be estimated without very detailed models of their sources.

Greisen, Zatsepin, and Kuzmin \cite{Greisen:1966jv,Zatsepin:1966jv} (GZK) realized that extragalactic protons can be attenuated by interactions with the background CMB photons of mean energy $E_\gamma \sim 0.23$ meV, via 
\be
p+\gamma\rightarrow \Delta^+ \rightarrow \pi^+ + n,
\ee
with threshold energy $E_p\simeq (m_\Delta^2 - m_p^2)/(4\,E_\gamma) \simeq 500$ EeV. Also heavier nuclei are attenuated by photo-disintegration of nuclei by CMB photons at similar energies.

As a result of GZK interactions, cosmogenic neutrinos originate from the secondary pions and neutrons from proton interaction in the same way than in sources, and then they add to the fluxes produced by the previously considered pion and neutron sources. Their contribution, first estimated by Berezinsky and Zatsepin in 1969 \cite{Beresinsky:1969qj}, has a maximum at energies of the order of EeV and depends on the CR mass composition and the assumed evolution with redshift of the sources.

\begin{figure}[t]
\begin{minipage}{.47\textwidth}
\includegraphics[width=1.\textwidth]{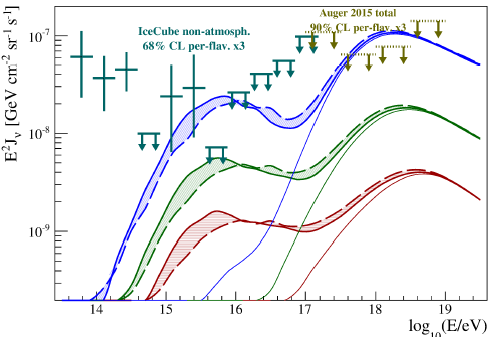}
\end{minipage}\hspace{2pc}%
\begin{minipage}{.47\textwidth}
\includegraphics[width=1.\textwidth]{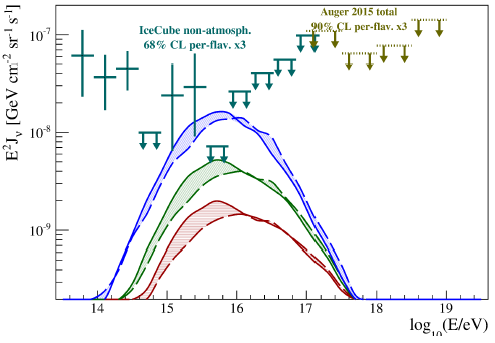}
\end{minipage} 
\caption{\label{gzknu}Fluxes of neutrinos expected at Earth for different cosmological evolution of sources (from bottom to top: no evolution, SFR, and AGN-like evolution) \cite{Aloisio:2015ega}. Left: pure proton composition, right: mixed composition.}
\end{figure}

In order to calculate the flux of GZK neutrinos, their interaction with cosmic matter is simulated by propagation codes, like SimProp or CRPropa. For example, the authors in \cite{Aloisio:2015ega} employ SimProp v2r2 with an injected spectrum of protons/nuclei, modeled according to the two mass composition indicated by CR data. Their results are shown in figure \ref{gzknu}. Left plot indicates that a pure proton composition, like in the {\it dip} model and TA data, with source evolution AGN-like, would result in a total neutrino flux in excess of the PAO and IceCube limits. On the other hand, PAO mass composition (right plot) would correspond to a somehow ``disappointing" universe with no observable neutrinos at high energy and only IceCube still competitive for detection at intermediate energy.

Then, the failure in detecting EeV neutrinos constrains proton-dominated models. GZK neutrino production could still proceed via the interaction of individual nucleons with background photons, but the threshold of the production would be increased to $E > A\, E_{GZK}$ for nuclei with mass number A. This scenario was investigated in \cite{Aloisio:2009sj} with the following assumptions: 1) proton composition in 1-3 EeV; 2) extragalactic origin; 3) rigidity dependent acceleration in sources, $E_{max} (Z) = Z\, E_p$ where $E_p$ is the maximum acceleration energy for protons. The results are very discouraging (motivating the adjective ``disappointing" used by the authors).
\begin{itemize}
\item
The maximum acceleration energy obtained for iron, $E_{max} \sim 100-300$ EeV, implies too small energy for nucleon ($2-5$ EeV) for satisfying the threshold condition for GZK production of neutrinos. Therefore, no cosmogenic neutrinos could be produced by protons on CMB, even if neutrinos of lower energies would come from interactions with Extra-galactic Background Light (EBL) photons.
\item
These $\nu$ from EBL photons are 6 times below the upper limit expected from IceCube after 5 years of observations, while $\nu$  from neutron decays even less than the EBL flux by two orders of magnitude.
\item
No sizable correlation of CRs with UHECR sources is observable, due to the deflection of nuclei in the galactic magnetic fields.
\end{itemize}

This brief discussion highlights the fact that a detection of neutrinos at EeV energies would help in sheding light on different mass composition scenarios.

\section{Neutrinos from sources of gravitational waves}

GWs, originated by the coherent, accelerated motion of astrophysically massive objects, are even more than neutrinos ideal universe messengers, because they travel almost unaltered (apart from cosmological red-shift) from sources to detectors.

We expect that a GW source should emit at the same time other more standard messengers. However, no significant counterpart was seen in the first historical GW observation from the binary BHs GW150914 \cite{Abbott:2016blz}. In any case, since the decay of a binary BH pair can be very slow, one does not expect surrounding matter at the time of coalescence. Thus the absence of others non gravitational messengers is not strange, as confirmed by the other GW events produced by binary BHs coalescences so far observed by LIGO. And, in fact, there is no theory of neutrino generation associated with BH-BH mergers, even if neutrinos could be emitted from BH accretion disk systems.

A more interesting situation, from the point of view of multi-messenger astronomy, is represented by the merger of binary NSs. And in fact, GW170817, the first observed merger of binary NSs on August 2017 \cite{TheLIGOScientific:2017qsa}, was also independently observed in electromagnetic radiation by Fermi, as a coincident GRB \cite{Monitor:2017mdv}. As for this kind of systems, the acceleration of particles by compact objects are still not well understood, and an observation of neutrinos could be the ``smoking gun" for hadronic processes and help in understanding the dissipation mechanisms in relativistic outflows.

I here consider two classes of sources that can emit both GWs and neutrinos.

{\bf GRBs}. According to the emergent picture of GRB, both short-duration (sGRB) and long-duration (lGRB) gamma ray bursts originate from binary systems involving at least one non-BH object. In the first case we have coalescing binaries involving at least one NS, in the second case the core-collapse of massive (few tens of solar masses) stars. It seems that the most promising neutrino production mechanism from sGRBs is connected to the Extended gamma Emission (EE), because its relatively low Lorentz factor results in high meson production efficiency \cite{Kimura:2017kan}. sGRBs have much smaller fluences than lGRBs, but we still have a chance connected to the late-time emission, which is less constrained by IceCube.

{\bf Magnetars}. Any stable remnant NS created by the merger of a binary system will be rotating rapidly, due to the large angular momentum. The remnant will also have ultra-strong internal magnetic fields ($>10^{15}-10^{16}$ G) and may acquire a similar strong external magnetic field ($>10^{13}-10^{15}$ G) \cite{Fang:2017tla}. The NS can power a relativistic wind with its rotational energy and the dense environment may collimate this wind into a bipolar jet with consequent long-lived X-ray emission. CRs accelerated in the magnetar nebula interact with ambient photons and baryons, producing pions and then TeV$-$PeV neutrinos and GeV photons.

\subsection{Neutrinos and sGRBs}

The light curves of sGRBs are a combination of different components: prompt emission, followed by EE, X-ray flares, and plateau emission (the classical afterglow, due to forward shocks propagating in the circumburst medium). Late time emission from the central engine is considered responsible for X and $\gamma$ rays, with energies comparable to the prompt emission \cite{Kimura:2017kan}.

\begin{figure}[t]
\includegraphics[width=.5\textwidth]{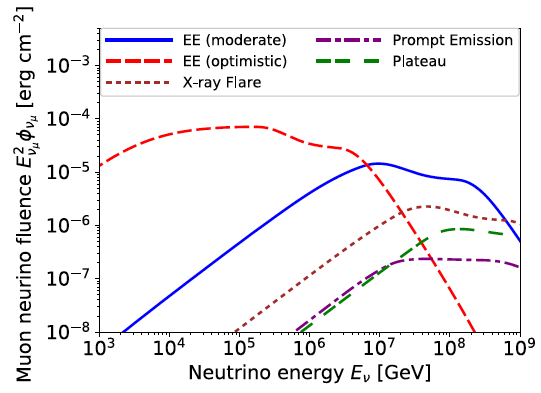}\hspace{2pc}%
\begin{minipage}[b]{.44\textwidth}\caption{\label{sGRBnu}Neutrino fluxes for different cases from \cite{Patrignani:2016xqp}: EE-mod, EE-opt, prompt emission, flare, and plateau for $d_L=300$ Mpc.}
\end{minipage}
\end{figure}

High-energy neutrino emission from GRBs has been studied \cite{Kimura:2017kan} with detailed numerical simulations. Based on the assumptions that 1) photon density is described by a broken power law, 2) CRs have the standard $E^{-2}$ power law, 3) cooling processes are represented by synchrotron cooling and photo-meson production and 4) the effect of oscillations eneters by tri-bimaximal mixing, they obtain the results shown in figure \ref{sGRBnu}. From the plot it is evident that EEs may be accompanied by more efficient production of high-energy neutrinos than the other components.

\subsection{Neutrinos and lGRBs}

Present models rely on the induced gravitational collapse paradigm \cite{Ruffini:2014tia}, based on the accretion process of the supernova ejecta of one component onto a NS binary companion. Photons are trapped within the accretion flow and thus cannot help in taking away the energy, but the $e^+ e^-$ pairs annihilate via weak interactions into equal numbers of neutrinos and antineutrinos, with energy of the order of $\sim 20$ MeV, which escape.

In particular, the peculiarity of these models stays in the fact that neutrino density near the NS surface is so high that the neutrino self-interaction potential, usually negligible in other very well known scenarios such as the Sun, becomes more relevant than the matter potential responsible for the Mikheyev-Smirnov-Wolfenstein effect. In \cite{Becerra:2017olp} it is shown that the initial $\phi_{\nu_e} : \phi_{\nu_x} = 7 : 3$ goes to $\phi_{\nu_e} : \phi_{\nu_x} = 5.5 : 4.5$ (for Normal Hierarchy) and $\phi_{\nu_e} : \phi_{\nu_x} = 6.2 : 3.8$  (for Inverted Hierarchy). These results can affect the $e^+ e^-$ production by neutrino pair annihilation, leading to measurable consequences on the GRB emission.

On the other side, TeV$-$PeV neutrinos can be produced from the interaction with the interstellar medium of CRs, e.g. protons, accelerated by the shock in the traditional collapsar scenario \cite{Woosley:1993wj}, where a GRB is produced in the fireball model with an ultra-relativistic jet.

\subsection{Neutrinos and magnetars}

As we have said, a remnant NS with strong magnetic field can transfer its rotational energy to a relativistic wind collimated into the dense environment created during the merger by the ejecta. The magnetar nebula is composed of $e^\pm$ and non-thermal photons at all wavelength (optical, UV, X-ray, $\gamma$-ray) produced by a cascade of high-energy photons coming from inverse Compton scattering of soft photons on background electrons and synchrotron emission in the nebula magnetic field. CRs with charge $Ze$ can be accelerated with very hard injection spectra $E^{-1}$ \cite{Fang:2017tla}.

During propagation, a CR interacts with $\gamma$ via PP and with ejecta baryons via HC and, at the same time, cools down by syncrotron radiation, which suppresses $\nu$ production.

Neutrinos are efficiently produced only if the pion or muon decay time is shorter than its cooling time. Since the ratio between the mean lifetimes of muons and pions is $\sim$ 100, muons are more likely to experience radiative cooling before decaying into secondary neutrinos.

\section{Multi-messenger astronomy and neutrinos}

\begin{figure}[t]
\includegraphics[width=.6\textwidth]{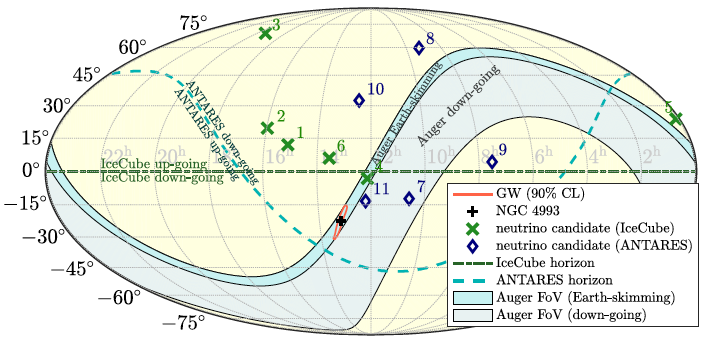}\hspace{2pc}%
\begin{minipage}[b]{.35\textwidth}\caption{\label{GW170817}Localizations and sensitive sky areas at the time of GW170817 in equatorial coordinates \cite{ANTARES:2017bia}.}
\end{minipage}
\end{figure}

The GW event of August 2017 was accompanied by the independent detection of a GRB by Fermi-LAT. The precise direction of GW170817 was above the ANTARES horizon at the detection time of the binary merger, see figure \ref{GW170817}. After considering also showers, no significant neutrino counterpart was found within a $\pm 500$ s window, nor in the subsequent 14 days \cite{ANTARES:2017bia}.
\begin{figure}[b]
\includegraphics[width=.37\textwidth]{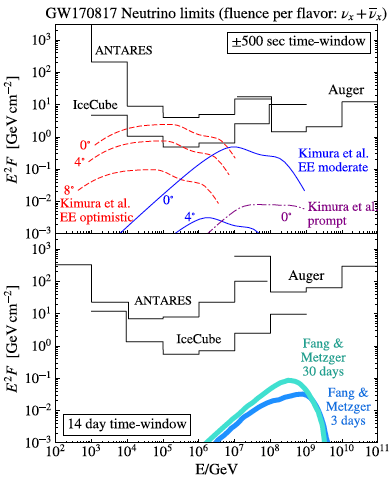}\hspace{2pc}%
\begin{minipage}[b]{.57\textwidth}\caption{\label{comparison}Comparison between experimental limits \cite{ANTARES:2017bia} and the theoretical prediction of \cite{Kimura:2017kan} and \cite{Fang:2017tla}.}
\end{minipage}
\end{figure}

Figure \ref{comparison} shows the comparison of the experimental limits with the theoretical prediction of two of the models of neutrino production considered in the previous sections, both of them scaled to a distance of 40 Mpc. In the upper panel, the EE model of \cite{Kimura:2017kan} was  applied for different inclinations of the jet-axis with respect to the line of view to Earth. No neutrino observation is consistent with off-axis sGRB. In the lower panel, the expected neutrino fluence from the magnetar model of \cite{Fang:2017tla} was considered. 

On September 2017 another multi-messenger event, this time involving neutrinos, has been detected, namely the neutrino event EHE170922A ($\sim$PeV) by IceCube \cite{IceCubeTXS} with a correspondent detections of gamma ray by Fermi-LAT \cite{FermiTXS} and MAGIC \cite{MAGICTXS} in a window of $\sim$5 days. The source was identified as the blazar (that is, AGN with its relativistic jet directed to Earth) TXS 0506+056 with red-shift $z = 0.3365\pm0.0010$.

Actually, blazars are considered potential neutrino emitters because their jets could offer suitable conditions to accelerate protons to the required high-energies. Blazar electromagnetic emission is generally attributed to radiating electrons (leptonic models) and has a low-energy component (first hump) arising from synchrotron radiation of electrons, and a high-energy one (second hump) typically attributed to inverse Compton scattering of ambient photons by the same electrons. Neutrinos are produced by CRs (hadronic models). In the theoretical model of blazar emission, a pure leptonic scenario would work but it would have no neutrinos, while a pure hadronic one would be hardly responsible for the second hump because synchrotron X-ray emission by secondary electrons would violate X-ray constraints.

\begin{figure}[t]
\begin{minipage}{.47\textwidth}
\includegraphics[width=1.\textwidth]{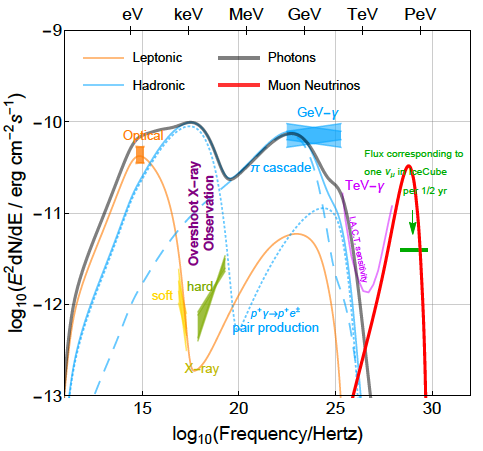}
\end{minipage}\hspace{2pc}%
\begin{minipage}{.47\textwidth}
\includegraphics[width=1.\textwidth]{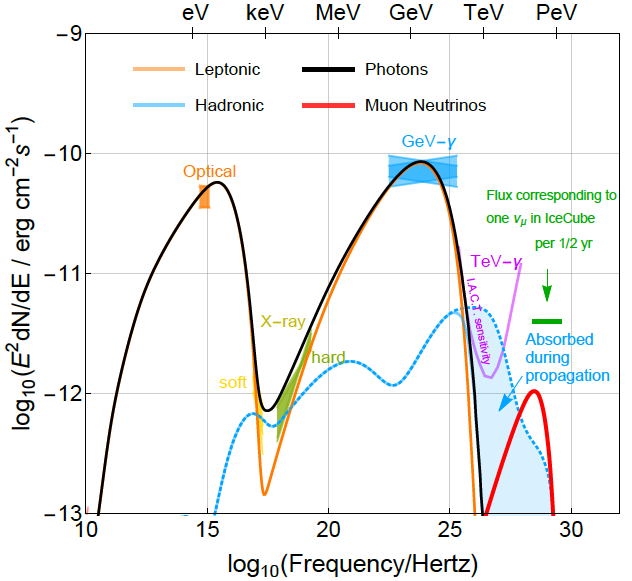}
\end{minipage} 
\caption{\label{TXS}Theoretical predictions for the spectral energy fluxes from TXS0506+056 flare for two hypothetical scenarios \cite{Gao:2018mnu}. Left: a simple hadronic model, in which the second hump comes from $\pi^0$ and $\pi^\pm$ decays. Right: a hybrid scenario with both leptonic and hadronic contributions.}
\end{figure}
In order to improve the performances of a pure hadronic model in describing the experimental data (figure \ref{TXS}, left plot), the authors in \cite{Gao:2018mnu} propose a hybrid model, with a core region in the jet moving at a different velocity, where the bulk of photon contribution is of leptonic origin and the hadronic contribution is the maximum allowed by the X-ray data (figure \ref{TXS}, right plot). As it is clear from the right plot, however, neutrino flux (red curve) is below the limit (green line) corresponding to one observed neutrino above 100 TeV in 180 days. They conclude that efficient neutrino production would require either a more compact production region during the flare or a somehow unphysical injected proton luminosity.

\section{Conclusions}

In this lectures I have reviewed the physics of cosmic neutrinos, with particular emphasis on their connection with other messengers arriving from the cosmo.

Neutrinos are traditionally considered, due to their very low interaction cross-sections, the ideal messengers from the universe but, while an advantage at the propagation stage, this characteristic becomes a handicap at the detection stage. Then, to be able to pursue the exciting goal of precise neutrino astronomy, new experimental techniques or upgrade of existing experiments are envisaged in the near future \cite{Pisanti:2017ncd}, like IceCube$-$Gen2 \cite{vanSanten:2017chb}, Auger-Prime \cite{Martello:2017pch}, KM3NeT ARCA and ORCA \cite{Piattelli:2015pmp,Brunner:2015ltd}, GVD in Lake Baikal \cite{Avrorin:2013uyc}, JEM$-$EUSO \cite{Takahashi:2009ng}.

Recent experimental detections show that multi-messenger combination of cosmic ray, photon, neutrino, and gravitational wave information is a very powerful tool in investigating the mechanisms for production and acceleration of particles in sources and their propagation in the universe.

These facts make us confident that multi-messenger strategy is fundamental and neutrino astronomy is an important part of it.

\end{document}